\documentclass[twocolumn]{aastex631}

\usepackage{amsmath}
\usepackage{cases}
\begin{document}

\title{Electron-scale Kelvin-Helmholtz instability in magnetized shear flows}

\author[0009-0009-7523-5887]{Yao Guo}
\affiliation{Key Laboratory for Laser Plasmas and Department of Physics and Astronomy, Collaborative Innovation Center of IFSA (CICIFSA), Shanghai Jiao Tong University, Shanghai 200240, People’s Republic of China}

\author[0000-0001-5738-5739]{Dong Wu}
\affiliation{Key Laboratory for Laser Plasmas and Department of Physics and Astronomy, Collaborative Innovation Center of IFSA (CICIFSA), Shanghai Jiao Tong University, Shanghai 200240, People’s Republic of China}

\author[0000-0001-7821-4808]{Jie Zhang}

\affiliation{Key Laboratory for Laser Plasmas and Department of Physics and Astronomy, Collaborative Innovation Center of IFSA (CICIFSA), Shanghai Jiao Tong University, Shanghai 200240, People’s Republic of China}
\affiliation{Tsung-Dao Lee Institute, Shanghai Jiao Tong University, Shanghai 201210, Peoples’s Republic of China}%

\correspondingauthor{Dong Wu}
\email{dwu.phys@sjtu.edu.cn}

\correspondingauthor{Jie Zhang}
\email{jzhang1@sjtu.edu.cn}

%% Note that the \and command from previous versions of AASTeX is now
%% depreciated in this version as it is no longer necessary. AASTeX 
%% automatically takes care of all commas and "and"s between authors names.

%% AASTeX 6.31 has the new \collaboration and \nocollaboration commands to
%% provide the collaboration status of a group of authors. These commands 
%% can be used either before or after the list of corresponding authors. The
%% argument for \collaboration is the collaboration identifier. Authors are
%% encouraged to surround collaboration identifiers with ()s. The 
%% \nocollaboration command takes no argument and exists to indicate that
%% the nearby authors are not part of surrounding collaborations.

%% Mark off the abstract in the ``abstract'' environment. 
\begin{abstract}

Electron-scale Kelvin-Helmholtz instabilities (ESKHI) are found in several astrophysical scenarios. Naturally ESKHI is subject to a background magnetic field, but an analytical dispersion relation and an accurate growth rate of ESKHI under this circumstance are long absent, as former MHD derivations are not applicable in the relativistic regime. We present a generalized dispersion relation of ESKHI in relativistic magnetized shear flows, with few assumptions. ESKHI linear growth rates in certain cases are numerically calculated. We conclude that the presence of an external magnetic field decreases the maximum instability growth rate in most cases, but can slightly increase it when the shear velocity is sufficiently high. Also, the external magnetic field results in a larger cutoff wavenumber of the unstable band and increases the wavenumber of the most unstable mode. PIC simulations are carried out to verify our conclusions, where we also observe the suppressing of kinetic DC magnetic field generation, resulting from electron gyration induced by the external magnetic field.

\end{abstract}

%% Keywords should appear after the \end{abstract} command. 
%% The AAS Journals now uses Unified Astronomy Thesaurus concepts:
%% https://astrothesaurus.org
%% You will be asked to selected these concepts during the submission process
%% but this old "keyword" functionality is maintained in case authors want
%% to include these concepts in their preprints.

\keywords{Relativistic jets(1390) --- Plasma jets(1263) --- Plasma physics(2089) --- Magnetic fields(994)}

%% From the front matter, we move on to the body of the paper.
%% Sections are demarcated by \section and \subsection, respectively.
%% Observe the use of the LaTeX \label
%% command after the \subsection to give a symbolic KEY to the
%% subsection for cross-referencing in a \ref command.
%% You can use LaTeX's \ref and \label commands to keep track of
%% cross-references to sections, equations, tables, and figures.
%% That way, if you change the order of any elements, LaTeX will
%% automatically renumber them.
%%
%% We recommend that authors also use the natbib \citep
%% and \citet commands to identify citations.  The citations are
%% tied to the reference list via symbolic KEYs. The KEY corresponds
%% to the KEY in the \bibitem in the reference list below. 

\section{Introduction} \label{sec:intro}
Shear instabilities play an important role in the evolution of astrophysical events by converting kinetic energy of the shear flow into thermal and electromagnetic energy and leading to formation of turbulence structures \citep{Weibel1959,Gruzinov1999Gamma,Medvedev1999,Silva_2003}, which will in turn accelerate particles to high velocities\ \citep{Rieger2019,Rieger2004,Ohira2013} and trigger non-thermal radiation emissions \citep{bottcher2007modeling}. As a result, understanding shear instabilities is the key to explaining the behavior of active galaxy nuclei (AGN) \citep{Colgate2001}, the mechanism of gamma ray bursts (GRB) \citep{Piran2005,Mirabel1999}, etc. 

Electron-scale Kelvin-Helmholtz instability (ESKHI) is a shear instability that takes place at the shear boundary where a gradient in velocity  is present. Despite the significance of shear instabilities, ESKHI was only recognized recently \citep{Gruzinov2008grb} and remains to be largely unknown in physics. Different from the classical KHI in neutral fluids or at an MHD scale \citep{helmholtz1868,Kelvin1871,chandrasekhar1961hydrodynamic}, ESKHI takes place on a much smaller timescale ($\tau \sim 1/\omega_{pe}$), where ions can hardly move due to their larger inertia. As a result, charge separation and generation of electromagnetic fields are characteristic during the development of ESKHI \citep{Grismayer2013dc}. Moreover, ESKHI can take place in relativistic shear flows ($\Delta v\sim c$), while classical fluid KHI is stable under a such condition \citep{Mandelker2016}. These make ESKHI a promising candidate to generate magnetic fields in the relativistic jets. 

ESKHI was first proposed by \citet{Gruzinov2008grb} in the limit of a cold and collisionless plasma, where he also derived the analytical dispersion relation of ESKHI growth rate for symmetrical shear flows. PIC simulations later confirmed the existence of ESKHI \citep{Alves2012large}, finding the generation of typical electron vortexes and magnetic field. It is noteworthy that PIC simulations also discovered the generation of a DC magnetic field (whose average along the streaming direction is not zero) in company with the AC magnetic field induced by ESKHI, while the former is not predicted by Gruzinov. The generation of DC magnetic fields is due to electron thermal diffusion or mixing induced by ESKHI across the shear interface \citep{Grismayer2013dc}, which is a kinetic phenomenon inevitable in the settings of ESKHI. A transverse instability labelled mushroom instability (MI) was also discovered in PIC simulations concerning the dynamics in the plane transverse to the velocity shear \citep{Liang2013Magnetic,Alves2015Transverse,Yao2020}. Shear flows consisting of electrons and positrons are also investigated \citep{Liang2013Magnetic,Liang2013Relativistic,Liang_2017}. Alves et al. investigated the mathematical details of ESKHI and numerically derived the dispersion relation in the presence of density contrasts or smooth velocity shears \citep{Alves2014electronshear}, which are both found to stabilize ESKHI.  \citet{miller2016relativistic} extended the theory of ESKHI to finite-temperature regimes by considering the pressure of electrons and derived a dispersion relation encompassing both ESKHI and MI. With regard to magnetic reconnection, secondary flux ropes generated by ESKHI has been observed in simulations \citep{Fermo2012Secondary} and already confirmed on the magnetopause by the observation of Magnetospheric Multiscale mission \citep{Zhong2018Evidence}.

In natural scenarios, ESKHI is often subject to an external magnetic field \citep{Niu2025,Jiang2025}. However, works mentioned above were all carried out in the absence of an external magnetic field. While the theory of fluid KHI has been extended to magnetized flows a long time ago \citep{chandrasekhar1961hydrodynamic,Dangelo1965}, the behavior of ESKHI in magnetized shear flows has been rather unclear. So far, the only theoretical considerations concerning this problem are presented by \citet{Che2023EKHI} and \citet{Tsiklauri_2024}. Both works are limited to incompressible plasmas and some kind of MHD assumptions, which are only valid for small shear velocities. Therefore, their conclusions cannot be directly applied in the relativistic regime, where ESKHI is expected to play a significant role \citep{Alves2014electronshear}. Simulations had reported clear discrepancies from their theory \citep{Tsiklauri_2024}. As Tsiklauri highlighted, a derivation of the dispersion relation without excessive assumptions is necessary. This forms part of the motivation behind our work.

In this paper, we will consider ESKHI under an external magnetic field by directly extending the works of  \citet{Gruzinov2008grb} and \citet{Alves2014electronshear}. This means that our work is carried out in the limit of cold and collisionless plasma. We adopt the relativistic two-fluid equations and avoid any form of MHD assumptions. The paper is organized as follows. In Sec.\ \ref{sec:intro}, we present a brief introduction to the background and subject of ESKHI. In Sec.\ \ref{sec:theo}, we present a theoretical analysis of ESKHI under an external magnetic field. A generalized dispersion relation (consisting of two coupled eigenvalue equations) of ESKHI growth rate is derived, and we numerically solve the growth rates for certain cases. In Sec.\ \ref{sec:sim}, we present and analyze the results of our PIC simulations to verify our theoretical analyses. In Sec.\ \ref{sec:conclusion}, we draw a conclusion and come up with some outlooks on future works. We use SI units throughout the paper, except for using Gauss (G) for the magnetic field when presenting simulation results in Sec.\ \ref{sec:sim}.

\section{Theoretical analyses} \label{sec:theo}

\subsection{Physical settings of the magnetized shear flows}

In this paper, we will stick with the relativistic fluid equations in the limit of cold and collisionless plasma. That is, we ignore the pressure term and the dissipation terms in the momentum equation:
\begin{align}
&\frac{\partial n}{\partial t}+\nabla\cdot(n\textbf{\textit{v}})=0, \label{eq:1}\\
&\left(\frac{\partial}{\partial t}+\textbf{\textit{v}}\cdot\nabla \right)\textbf{\textit{p}}+e\left(\textbf{\textit{E}}+\frac{\textbf{\textit{p}}}{\gamma m}\times \textbf{\textit{B}}\right)=0,    \label{eq:2}
\end{align}
where the momentum $\textbf{\textit{p}}=\gamma m \textbf{\textit{v}}$ and the Lorentz factor $\gamma=1/\sqrt{1-v^2/c^2}$. These equations are coupled with the Maxwell equations
\begin{align}
&\nabla\times \textbf{\textit{E}}=-\frac{\partial \textbf{\textit{B}}}{\partial t},\label{eq:3}\\
&c^2 \nabla\times \textbf{\textit{B}}=-\frac{1}{\epsilon_0}\textbf{\textit{J}}+\frac{\partial \textbf{\textit{E}}}{\partial t},\label{eq:4}
\end{align}
and the electric current is
\begin{equation}
\textbf{\textit{J}}=-en_e\textbf{\textit{v}}_e+en_0\textbf{\textit{v}}_i. \label{eq:5}
\end{equation}

ESKHI can be triggered when a velocity shear is present. Without loss of generality, we consider an electron-ion (protons, to be specific) shear flow with an interface located at $x=0$ and a non-uniform initial velocity distribution $\textbf{\textit{v}}_{e0}=\textbf{\textit{v}}_{i}=(0,v_0(x),0)$. The ions are considered to be free-streaming and unperturbed throughout the timescale. The initial number density distribution is $n_{e0}(x)=n_i(x)$.  An external magnetic field $\textbf{\textit{B}}_0=(0,B_0(x),0)$ in the streaming direction of the shear flow is imposed. In the cold limit, a dynamical equilibrium requires $ \textbf{\textit{E}}_0=0$, given $ \textbf{\textit{v}}_{e0}\times  \textbf{\textit{B}}_0=0$. No net current $\textbf{\textit{J}}_{0}$ results in a uniform magnetic field $B_0(x)=\text{const.}$  This setting is in hydro equilibrium but not in Vlasov equilibrium, which will give rise to a DC magnetic field \citep{Grismayer2013dc}. For physical interest related with magnetic reconnection, we can also consider a step function of $B_0(x)$ with a jump at $x=0$. This results in an infinitely thin current sheet in $Oz$ direction at the interface $x=0$, but does not affect the linear behavior outside the interface.

We consider perturbations in the streaming direction $Oy$, and all the perturbed variables are in the form of $f(x)e^{i(ky-\omega t)}$. In principle, we can also consider perturbation in the transverse direction $Oz$ to contain ESKHI, MI \citep{Alves2015Transverse}, and their coupling in a once-for-all dispersion relation. However, it will add more complexity to our already complicated enough problem. Hence, for demonstration we will stick with perturbations in the $Oy$ direction and assume them to be uniform in the $Oz$ direction. It can be seen as a special case for $k_z=0$ in perturbations with the form $f(x)e^{i(k_yy+k_zz-\omega t)}$. 

\subsection{Review on former efforts}

Before we start, we would like to have a brief review on former works which form the basis of our work. \citet{Gruzinov2008grb} firstly derived the dispersion relation of ESKHI in an unmagnetized shear flow. The work was in the limit of cold and collision plasma, and the equations used are just our Eqs.\ \ref{eq:1}$\sim$\ref{eq:5} (see \citet{Alves2014electronshear} for details). The generalized eigenequation he derived is
\begin{equation}
    \left[\frac{\omega_{p}^2-\Omega^2}{A(\omega,k)} E_y'   \right]'+\frac{1}{c^2}\left(\frac{\omega_{p}^2}{\Omega^2}-1\right)E_y=0,
\end{equation}
where $\Omega=\omega-kv_0$, $A(\omega,k)=\Omega^2\left(\omega^2-k^2c^2-\gamma_0^2\omega_{p}^2\right)$, $\gamma_0=1/\sqrt{1-v_0^2/c^2}$ is the Lorentz factor of the shear flow, and $\omega_p=\sqrt{n_{e0} e^2/m_e\epsilon_0\gamma_0^3}$ is the relativistic electron plasma frequency. Here we note that in this paper the ion plasma frequency $\omega_{pi}$ is irrelevant and all appearances of $\omega_{p}$ refer to $\omega_{pe}$. For a symmetrical shear flow ($n_{e0}(x)=n_0, v_0(x)=\text{sgn}(x)V_0$), the dispersion relation of the growth rate $\sigma=\text{Im}(\omega)$ is
\begin{equation}
    \sigma^2=\frac{\omega_p^2}{2}\left(\sqrt{1+8\frac{k^2V_0^2}{\omega_p^2}}-1-2\frac{k^2V_0^2}{\omega_p^2}\right), \label{eq:gru}
\end{equation}
for $0\leq kV_0\leq\omega_p$. Perturbations with $kV_0>\omega_p$ are all stable.

\citet{Che2023EKHI} was the first analytical study on magnetized ESKHI, where they studied a semi-uniform shear flow with $n_{e0}(x\gtrless0)=n_{1,2}$, $v_0(x\gtrless0)=V_{1,2}$ and $B_0(x\gtrless0)=B_{1,2}$ respectively for the upper and lower half of the plane. Their motivation is to extend MHD KHI to the electron scale, so there is no wonder they derived a dispersion relation similar to that of MHD KHI \citep{chandrasekhar1961hydrodynamic}:
\begin{equation}
    \sigma^2=\frac{n_1n_2\Delta V^2-(n_1+n_2)(n_1 v_{A1}^2+n_2v_{A2}^2)}{(n_1+n_2)^2}k^2, \label{eq:che}
\end{equation}
where $\Delta V=V_1-V_2$, $v_{Aj}=B_j/\sqrt{\mu_0m_en_{j}}$ is the electron Alfven velocity in each half of the plane. For $n_1=n_2=n_0$, this equation predicts a threshold magnetic field $B_{\text{MHD}}=\sqrt{\mu_0m_en_0}V_0$ beyond which all perturbations are stable. Also, the direction of the magnetic field does not affect the growth rate as it is enveloped in $v_{Aj}^2$. The latter work of \cite{Tsiklauri_2024} derived the same dispersion relation with slightly different assumptions. One can see that Eq.\ \ref{eq:che} cannot be reduced to Eq.\ \ref{eq:gru} in absence of external magnetic fields, i.e. $v_{Ai}=0$, unless in the limit of infinitely small $k\Delta V$. In fact, Eq.\ \ref{eq:che} predicts a divergent growth rate for infinitely large wavenumber $k$ while Eq.\ \ref{eq:gru} explicitly gives a cutoff wavenumber above which all perturbations are stable. The discrepancy is not surprising, as Che and Zank adopted the incompressible condition $\nabla\cdot \textbf{\textit{v}}=0$ and no vorticity condition $\nabla \times \textbf{\textit{v}}=0$, and both assumptions are valid for small shear velocities. Moreover, Che and Zank chose to neglect the induced electric field in Eq.\ \ref{eq:4}. This assumption is also valid for sufficiently small shear velocities, where $\partial \textbf{\textit{E}}/\partial t\sim \sigma\Delta \textbf{\textit{E}} \propto kV_0 \Delta \textbf{\textit{E}}$. But for sub-relativistic and relativistic flows, the growth rate $\sigma$ is on the order of $\omega_p$, hence $\partial \textbf{\textit{E}}/\partial t$ is sufficiently large and cannot be neglected. In this case, the MHD frozen-in condition $\textbf{\textit{E}}+\textbf{\textit{v}}\times\textbf{\textit{B}}=0$ adopted by \citet{Tsiklauri_2024} is also not suitable. Thus, we conclude that Eq.\ \ref{eq:che} is only applicable in the non-relativistic limit, and a dispersion relation in the sub-relativistic and relativistic regime is to be determined without excessive assumptions. We will focus on this problem in the preceding subsection.

\subsection{Mathematical formulation for the dispersion relation}

Now we linearize Eqs.\ \ref{eq:1}$\sim$\ref{eq:5} with the perturbations in the form of $f(x)e^{i(ky-\omega t)}$. The perturbed variables are $n_1,\ \textbf{\textit{v}}_{e1}=(v_x,v_y,v_z),\ \textbf{\textit{B}}_1=(B_z,B_y,B_z),\ \textbf{\textit{E}}_1=(E_x,E_y,E_z)$, and $\textbf{\textit{J}}_1=(J_x,J_y,J_z)$. The linearized form of Eqs. \ref{eq:1} and \ref{eq:2} is
\begin{align}
&n_1=\frac{-i}{\Omega}\left[\frac{\partial }{\partial x}(v_xn_{e0})+ikn_{e0}v_y\right],\label{eq:n1}\\
&    v_x=\frac{e}{\gamma_0 m_e}\frac{-i}{\Omega}(E_x+v_0B_z-v_zB_0), \label{eq:vx} \\    
&    v_y=\frac{e}{\gamma_0^3 m_e}\frac{-i}{\Omega}\left[E_y+\frac{m_e}{e}v_x\frac{\partial}{\partial x}(\gamma_0 v_0)\right],\label{eq:vy} \\
&    v_z=\frac{e}{\gamma_0 m_e}\frac{-i}{\Omega}(E_z-v_0B_z+v_xB_0).  \label{eq:vz}  
\end{align}
One can see here that the existence of the external magnetic field $B_0$ couples $v_x$ and $v_z$ linearly. We can solve Eqs.\ \ref{eq:vx} and \ref{eq:vz} to decouple them:
\begin{align}
&    v_x=-\frac{e}{\gamma_0 m_e}\frac{\omega_c (-E_z+B_zv_0)+i\Omega(E_x+B_zv_0)}{\Omega^2-\omega_c^2},\label{eq:vx1} \\
& v_z=-\frac{e}{\gamma_0 m_e}\frac{\omega_c(E_x+B_zv_0)+i\Omega(E_z-B_zv_0)}{\Omega^2-\omega_c^2},\label{eq:vz1}
    \end{align}
where $\omega_c=eB_0/m_e\gamma_0$ is the relativistic electron cyclotron frequency. By this definition, $\omega_c$ can be negative with a negative $B_0$. The linearization of Eq. \ref{eq:5} is 
\begin{equation}
    \textbf{\textit{J}}=-e(n_1\textbf{\textit{v}}_{e0}+n_{e0}\textbf{\textit{v}}_{e1}). \label{eq:J1}
\end{equation}
Substituting Eqs.\ \ref{eq:n1}, \ref{eq:vy}, \ref{eq:vx1}, and \ref{eq:vz1} into Eq.\ \ref{eq:J1}, we can derive
\begin{align}
  &  J_x=\frac{e^2n_{e0}}{\gamma_0m_e}\frac{\omega_c (-E_z+B_zv_0)+i\Omega(E_x+B_zv_0)}{\Omega^2-\omega_c^2 },\label{eq:Jx}\\
  &  J_y=\frac{e^2n_{e0}}{\gamma_0^3 m_e}\frac{i\omega}{\Omega^2}E_y+ie\frac{\partial}{\partial x}\left(\frac{v_0v_xn_0}{\Omega}\right),\label{eq:Jy} \\
  &  J_z=\frac{e^2n_{e0}}{\gamma_0m_e}\frac{\omega_c (E_x+B_zv_0)+i\Omega(E_z-B_zv_0)}{\Omega^2-\omega_c^2 }.\label{eq:Jz}
\end{align}
Then we linearize the Maxwell equations Eqs.\ \ref{eq:3} and \ref{eq:4} and couple them together to derive
\begin{align}
&B_z=\frac{k}{\omega}E_z, \label{eq:Bx}\\
&B_z=-\frac{i}{\omega}\frac{\partial E_y}{\partial x}-\frac{k}{\omega}E_x,\label{eq:Bz}
\end{align}
and
\begin{equation}
\nabla\times(\nabla\times \textbf{\textit{E}}_1)=-\frac{1}{c^2}\left(\frac{1}{\epsilon_0}\frac{\partial \textbf{\textit{J}}_1}{\partial t}+\frac{\partial^2 \textbf{\textit{E}}_1}{\partial t^2}\right),\label{eq:ME}
\end{equation}
whose $x$ component is
\begin{equation}
ik\frac{\partial E_y}{\partial x}=i\omega \mu_0 J_x+\left(\frac{\omega^2}{c^2}-k^2\right)E_x.\label{eq:MEx}
\end{equation}
Note that $J_x$ contains only the first order terms of $E_x$, so with Eqs.\ \ref{eq:Jx}, \ref{eq:Bx}, \ref{eq:Bz}, and \ref{eq:MEx}  we can solve $E_x$ out, in terms of only $E_y$ and $E_z$:
\begin{eqnarray}
E_x=&&\frac{i\left[kc^2\left(\Omega^2-\omega_c^2 \right)-\gamma_0^2\omega_p^2\Omega v_0\right]}{D(\omega,k)} \frac{\partial E_y}{\partial x}\nonumber \\
&&+\frac{i\gamma_0^2\omega_p^2\omega_c\Omega}{D(\omega,k)} E_z, \label{eq:Ex}
\end{eqnarray}
where $D(\omega,k)=A(\omega,k)-\omega_c^2(\omega^2-k^2c^2)=\Omega^2\left(\omega^2-k^2c^2-\gamma_0^2\omega_p^2\right)-\omega_c^2(\omega^2-k^2c^2)$. The $y$ and $z$ components of Eq.\ \ref{eq:ME} are
\begin{align}
ik\frac{\partial E_x}{\partial x}-\frac{\partial^2 E_y}{\partial x^2}=i\omega \mu_0 J_y+\frac{\omega^2}{c^2}E_y,\label{eq:MEy} \\
-\frac{\partial^2 E_z}{\partial x^2}=i\omega \mu_0 J_z+\left(\frac{\omega^2}{c^2}-k^2\right)E_z. \label{eq:MEz}   
\end{align}
Substitute Eqs.\ \ref{eq:vx1}, \ref{eq:Jy}, \ref{eq:Jz}, \ref{eq:Bx},  \ref{eq:Bz}, and \ref{eq:Ex} into Eqs.\ \ref{eq:MEy} and \ref{eq:MEz}, and after some algebra we can derive two coupled differential equations with only two perturbed variables $E_y,E_z$. These two equations are just our generalized eigenvalue equations:
\begin{widetext}
\begin{numcases}{}
   \left[\frac{\omega^2(\omega_c^2+\omega_p^2-\Omega^2)}{D(\omega,k)}E_y'\right]'+\frac{\omega^2}{c^2}\left(\frac{\omega_p^2}{\Omega^2}-1 \right)E_y  - 
   \left[\frac{\omega\omega_c\gamma_0^2\omega_p^2\left(k-\omega v_0/c^2\right)}{D(\omega,k)}E_z\right]'=0 ,\label{eq:dis1} \\ 
    E_z''+\frac{1}{c^2}\frac{\Omega^2(\omega^2-k^2c^2-\gamma_0^2\omega_p^2)^2-\omega_c^2(\omega^2-k^2c^2)^2}{D(\omega,k)}E_z-\frac{\omega\omega_c\gamma_0^2\omega_p^2\left(k-\omega v_0/c^2\right)}{D(\omega,k)}E_y'=0,  \label{eq:dis2}
\end{numcases}
where the ``$'$" notation represents the partial derivative with respect to the $x$ coordinate. In the case of a discontinuous shear interface at $x=0$, we can integrate Eqs.\ \ref{eq:dis1} and \ref{eq:dis2} across the interface to derive two boundary conditions
\begin{numcases}{}
 \left[\frac{\omega^2(\omega_c^2+\omega_p^2-\Omega^2)}{D(\omega,k)}E_y'\right]\bigg|^{0_+}_{0_-}- \left[\frac{\omega\omega_c\gamma_0^2\omega_p^2\left(k-\omega v_0/c^2\right)}{D(\omega,k)}E_z\right] \bigg|^{0_+}_{0_-}=0 ,\label{eq:dis10} \\ 
    E_z'(0_+)=E_z'(0_-),  \label{eq:dis20}
\end{numcases}
\end{widetext}
Equations\ \ref{eq:dis1} and \ref{eq:dis2} are generic and not confined to specific profiles of the zeroth-order variables. In principle, one can assign any profile of the electron density $n_{e0}(x)$ and shear velocity $v_0(x)$ along with any boundary conditions and use Eqs.\ \ref{eq:dis1} and \ref{eq:dis2} to check the stability of the system.

In the absence of the external magnetic field, i.e. $\omega_c=0$, one can easily verify that $E_y$ decouples from $E_z$ and Eq.\ \ref{eq:dis1} naturally reduces to the dispersion relation of unmagnetized ESKHI, Eq.\ \ref{eq:gru}, derived by  \citet{Gruzinov2008grb}, while Eq.\ \ref{eq:dis2} actually reduces to the dispersion relation of an electromagnetic wave, $[\omega^2-(k_x^2+k_y^2)c^2-\gamma_0^2\omega_p^2]E_z=0$. This indicates that the effect of the external magnetic field is to couple the ESKHI mode with electromagnetic modes with components in the transverse direction.

\subsection{Numerical calculated growth rates} \label{sec:num}
In the simplest case of a symmetrical shear flow, this set of equations can be solved analytically in principle. However, the form of $E_z$ and $E_y$ in the solution to Eqs. \ref{eq:dis1} and \ref{eq:dis2} is indeed enormous, making it unrealistic to derive an analytical expression of the growth rate. 

\begin{figure}[h]
\centering
\includegraphics[width=0.98 \columnwidth]{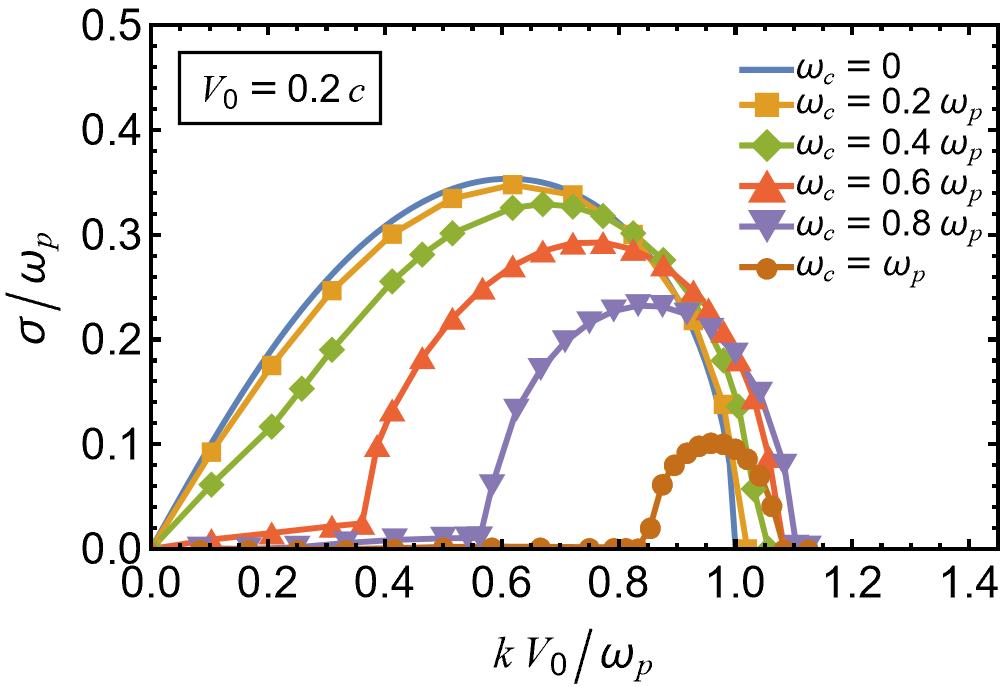}
\caption{The dispersion relation of the ESKHI growth rate with $v_0=0.2c \text{sgn}(x)$ under  uniform external magnetic fields $B_0(x)=B_E$ corresponding to $\omega_c/\omega_p=eB_E/m_e\gamma_0\omega_p=$0.2 (yellow, squares), 0.4 (green, diamonds), 0.6 (red, upward triangles), 0.8 (purple, downward triangles), and 1 (brown, circles). Each marker represents a calculated point. The theoretical curve in absence of external magnetic field (blue) is also shown for comparison. For each wavenumber, we choose only the eigenfrequency with the largest imaginary part.}
\label{fig:disu}
\end{figure}

\begin{figure}[h]
\centering
\includegraphics[width=0.98 \columnwidth]{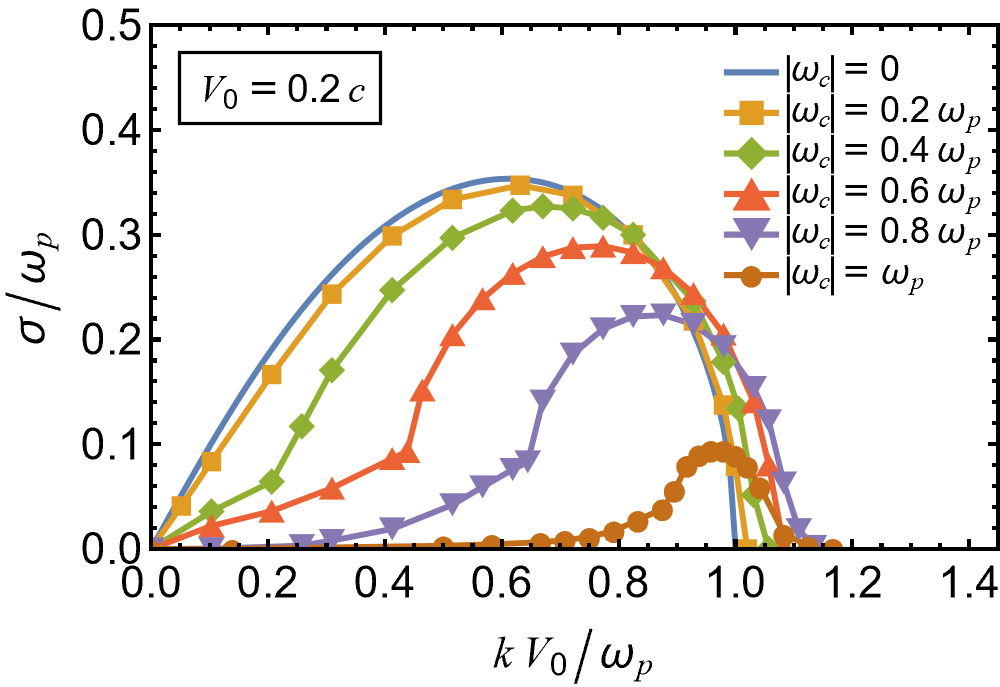}
\caption{The same plot as Fig.\ \ref{fig:disu}, but under anti-parallel external magnetic fields $B_0(x)=B_E\text{sgn}(x)$.}
\label{fig:disn}
\end{figure}

Here we present the numerical solutions of the dispersion relation for some simple cases. The detail of the numerical scheme is introduced in Appendix\ \ref{sec:app}. Figure\ \ref{fig:disu} shows the dispersion relation of ESKHI in an infinite shear flow under uniform magnetic fields of different magnitudes and Fig.\ \ref{fig:disn} for anti-parallel magnetic fields $B_0(x)=B_E\text{sgn}(x)$, with a symmetrical velocity profile $v_0=V_0\text{sgn}(x)=0.2c\text{sgn}(x)$ in both cases.
At some wavenumbers, multiple eigenvalues exist and we only plot the eigenfrequency with the largest imaginary part. In each case, the existence of the external magnetic field decreases the maximum growth rate of ESKHI, while increasing the cutoff wavenumber of the unstable band. This is because on one hand, the magnetic tension plays a suppressing role for perturbations with small $kV_0$ \citep{chandrasekhar1961hydrodynamic}. On the other hand, for perturbations with $kV_0$ close to $\omega_p$, which could not have extracted sufficient energy from the electrons in absence of the external magnetic field, $\omega_c$ effectively enlarges $\omega_p$, allowing the perturbations to be amplified. As a result, the wavenumber of the most unstable mode increases as $B_E$ gets larger. 

Remember that Eq.\ \ref{eq:che} proposed a threshold magnetic field $B_{\text{MHD}}=\sqrt{\mu_0m_en_0}V_0$, which corresponds to a $\omega_c=\beta_0 \omega_p=\omega_pV_0/c$ in the non-relativistic limit. In the relativistic regime, the latter condition corresponds to a $B_{M}=\sqrt{\mu_0m_en_0/\gamma_0}V_0$. From Figs.\ \ref{fig:disu} and \ref{fig:disn}, it is clear that in both cases the unstable modes of ESKHI survive beyond $B_M$, which corresponds to $\omega_c=0.2\omega_p$ in our cases. Up to $\omega_c=1.2\omega_p$, perturbations of all wavenumbers are stabilized in both cases. Although we don't know the actual form of the threshold magnetic field, it seems to be closely related with $\omega_p$ since the growth rates decrease rapidly when $\omega_c$ approaches and surpasses $\omega_p$. Physically, when the electron cyclotron frequency is larger than the electron plasma frequency, any electrons with velocity across the interface will be re-directed before it gets amplified sufficiently. Also, the threshold magnetic field is different for different $kV_0$, as for $\omega_c=\omega_p$ perturbations with smaller wavenumbers are already stable while those with larger wavenumbers are not.

One can also see from Figs.\ \ref{fig:disu} and \ref{fig:disn} that the curves of the dispersion relations are not overall smooth, while some turning points clearly separate parts with different trends of the curve. It actually marks different behaviors of ESKHI over the wavenumber band. For small wavenumbers and relatively large magnetic fields (e.g. $kV_0/\omega_p\lesssim0.35,\omega_c=0.6\omega_p$ in the uniform case), we find that the eigenfrequency $\omega$ is essentially complex, which corresponds to a wave both growing and travelling, resulting from coupling to electromagnetic waves. For larger wavenumbers (e.g. $0.35\lesssim kV_0/\omega_p\lesssim1.1,\omega_c=0.6\omega_p$ in the uniform case) the frequency is purely imaginary, which corresponds to a purely growing wave. The mathematical and physical mechanism causing this transition is not clear so far.

Also, while Eq.\ \ref{eq:che} does not distinguish the cases of uniform and anti-parallel magnetic fields, our dispersion relations are clearly distinct for the two different cases, as can be seen from Figs.\ \ref{fig:disu} and \ref{fig:disn}. Under the same magnitude of external magnetic fields $B_E$, the growth rate of the travelling part is obviously larger in the anti-parallel case than in the uniform case. This results from the interplay between the velocity shear and the anti-parallel magnetic field configuration, which itself can trigger tearing instabilities \citep{Coppi1966,Pritchett1996} in other contexts. However, we would like to note that our Eqs.\ \ref{eq:dis1} and \ref{eq:dis2} with $V_0=0$ and merely $B_0(x)=B_E\text{sgn}(x)$ do not admit an unstable solution. It is not surprising, as in MHD framework tearing instabilities are only unstable when considering resistive effects, which are not reflected in our cold fluid equations.

\begin{figure}[h]
\centering
\includegraphics[width=0.98 \columnwidth]{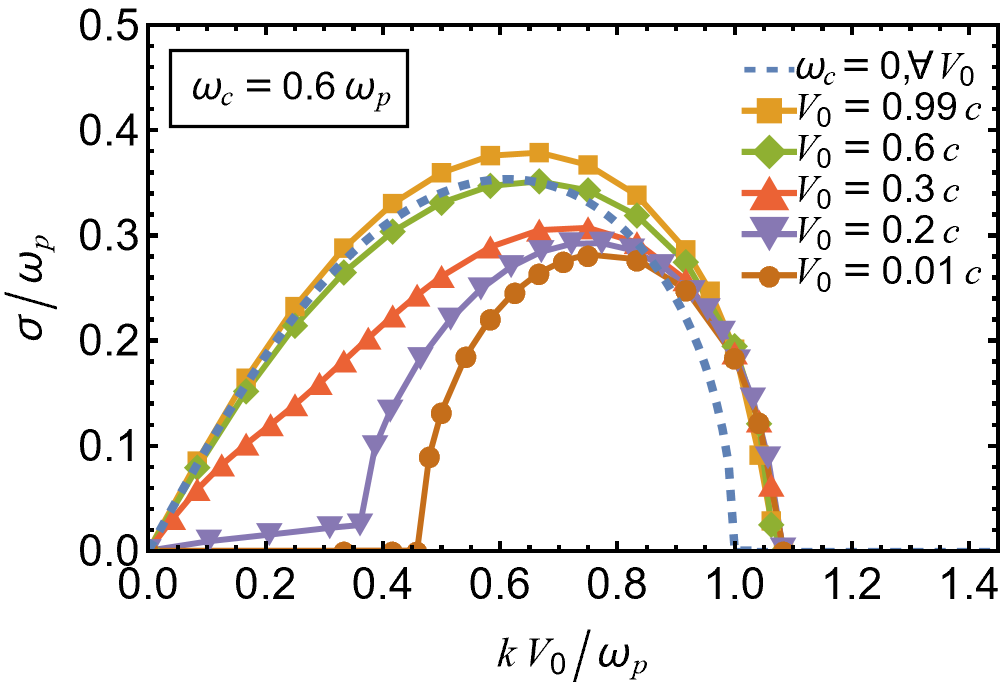}
\caption{The dispersion relation of the ESKHI growth rate under a uniform magnetic field with $\omega_c=0.6\omega_p$ and different shear velocities $V_0/c=$0.99 (yellow, squares), 0.6 (green, diamonds), 0.3 (red, upward triangles), 0.2 (purple, downward triangles), and 0.01 (brown, circles). The theoretical curve in absence of external magnetic field (blue, dashed) is also shown for comparison, whose profile in the normalized variables does not depend on specific values of $V_0$.}
\label{fig:disv}
\end{figure}
In the absence of the external magnetic field, the normalized growth rate $\sigma/\omega_p$ only depends on the value of $kV_0/\omega_p$, as can be seen in the dispersion relation Eq.\ \ref{eq:gru}. However, we would like to emphasize that, in the presence of the external magnetic field, occurrence of $\omega_p$ breaks the self-similarity of the dispersion relation. Also, $\gamma_0$ seems to be relevant in Eq.\ \ref{eq:dis1} and \ref{eq:dis2}, while it eventually vanishes in Eq.\ \ref{eq:gru}. As a result, $\sigma/\omega_p$ will depend on the specific value of $V_0/c$. Figure\ \ref{fig:disv} shows the dispersion relation of ESKHI with $\omega_c=0.6\omega_p$ but different shear velocities. It is clear that the shear flows with smaller velocities are more affected by the external magnetic field, while for larger shear velocities the growth rate at lower wavenumbers is almost the same as that in absence of the magnetic field. At high wavenumber end, the profile of the normalized growth rate seems to be the same for all shear velocities. It indicates a normalized cutoff wavenumber $k_cV_0/\omega_p$ dependent on only $\omega_p$ and $\omega_c$ but not specific values of $V_0$. For $V_0=0.0001c$ (not plotted), we derived a profile very close to that of $V_0=0.01c$, indicating an asymptotic behavior for small $V_0$. Interestingly, for high shear velocities $V_0=0.6c$ and $0.99c$, we find a slight increase in the maximum growth rate with the presence of the external magnetic field, which was never predicted in former works. It is a consequence of little suppressing at the low wavenumber end combined with obvious activation at the high wavenumber end. We can imagine that for sufficiently large shear velocities, external magnetic fields below a critical magnitude are destabilizing for ESKHI, while stabilizing above the critical magnitude. As this increase is not so evident, we will not conduct further calculations on it. Ones interested may test different $\omega_c$ with a large $V_0$ to see the trend.

Finally, we would like to note that we neglected the effect of ion motions. This will only lead to a difference in the order of $m_e/m_i$ in our settings \citep{Alves2014electronshear}. However, in an electron-positron plasma or an electron-positron-ion plasma, this will lead to a correction factor up to $\sqrt{2}$ to the growth rate. Anyway, it does not affect the physical essence of our work, as the ions merely obey the same equations as the electrons do, except for the change in masses (and charges).

\section{PIC simulations} \label{sec:sim}
In order to verify our analytical results, we carry out 2D simulations in the $xOy$ plane with a full kinetic PIC code LAPINS \citep{Wu2019}. The velocity space is 3D. The initial density profile is $n_e=n_i=10^{15}\ \text{m}^{-3}$. The simulation area is $L_x\times L_y=3.2\ \text{m}\times9.6\ \text{m}$, divided into $800\times 2400$ grids. Each grid scales as $4\ \text{mm}\times 4\ \text{mm}$. The length in the $Oy$ direction is intentionally set long enough to resolve dozens of wavelengths of the unstable modes, which can facilitate our Fourier analyses later. The shear velocity is $V_0=0.2c$, and the profile is the ``sandwich''  configuration often used in KHI-relevant simulations:
    \begin{equation}
    v_{y}(x)=\begin{cases}
    V_0,& \text{if } x < 0.8\ \text{m}, \\
    -V_0,& \text{if } 0.8\ \text{m}\leq x \leq 2.4\ \text{m},\\
    V_0,& \text{if } x > 2.4\ \text{m}.
    \end{cases}
    \label{eq:velocity}
\end{equation}
Periodic boundary conditions are implemented in both directions. The Lorentz factor is $\gamma_0=1/\sqrt{1-(V_0/c)^2}\approx1.02$, resulting in a corresponding electron plasma frequency  $\omega_{p}\approx 1.73\times10^9\ \text{s}^{-1}$ and an electron skin depth $c/\omega_p\approx0.17$ m. The initial temperature profile is $T_e=T_i=0.5$ eV, which results in a thermal velocity of $v_{Te}=0.0014 \text{c}\ll V_0 $ and satisfies the cold plasma assumption. The uniform external magnetic field is pre-imposed in the $Oy$ direction, with different magnitude of $B_0= 0,3,4,5 B_M$ in different runs, which corresponds to $\omega_c=0,0.6,0.8,1\omega_p$  respectively. The settings of each run just correspond to the cases investigated in Fig.\ \ref{fig:disu}, which facilitates our verification of the analytical results.

\begin{figure}[h]
\centering
\includegraphics[width=0.98 \columnwidth]{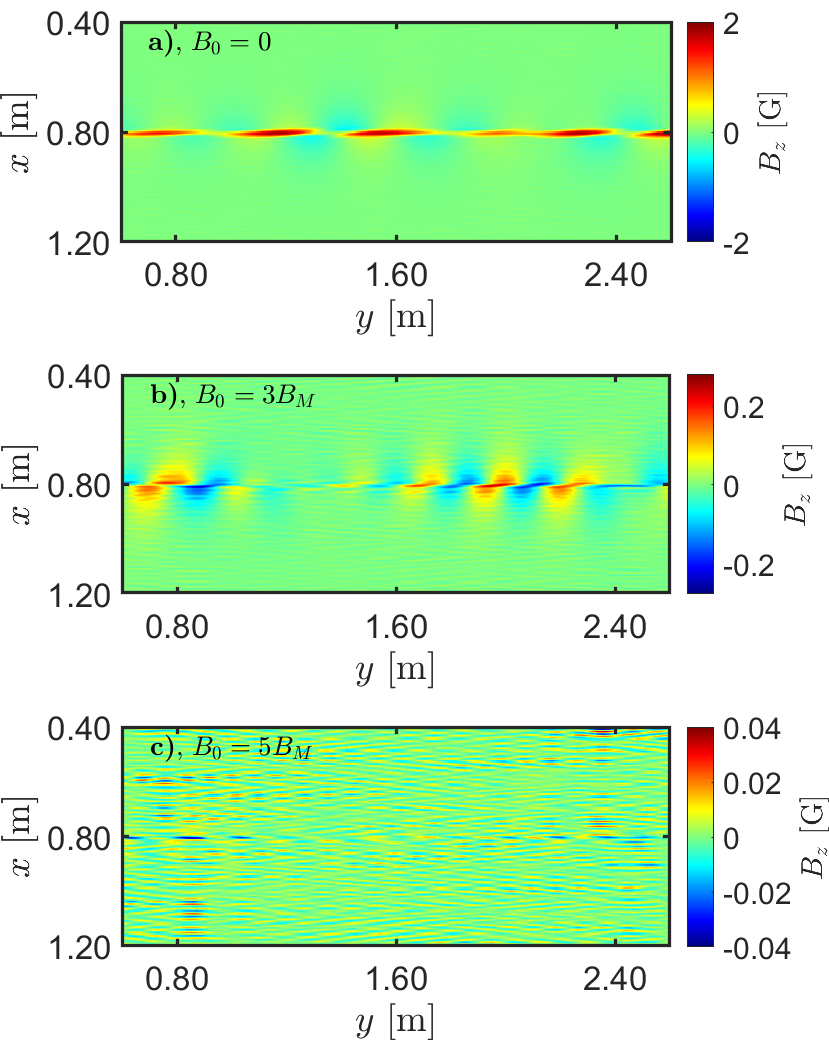}
\caption{The transverse magnetic field $B_z$ around the shear interface in runs of \textbf{a)}, $B_0=0$, \textbf{b)}, $B_0=3B_M$, and \textbf{c)}, $B_0=5B_M$ at $t=28.5\omega_p^{-1}$. Each figure is restricted in the same area, which is $1/24$ of the simulation domain. A color-bar is assigned to each panel respectively.}
\label{fig:Bx}
\end{figure}

In each simulation, we observe the generation of ESKHI modes along with the DC magnetic fields. Figure\ \ref{fig:Bx} shows the transverse magnetic field $B_z$ extracted at the same simulation time $t=28.5\omega_p^{-1}$ from three runs of $B_0=0,3,5B_M$ respectively. The characteristics of $B_z$ are qualitatively consistent with our analytical results. It is clear that as the magnitude of the external magnetic field $B_0$ increases, the magnitude of the generated transverse magnetic field $B_z$ decreases. For $B_0=5B_M$, the plotted time is not long enough for the unstable mode to grow up and dominate. We can also see that in Fig.\ \ref{fig:Bx}b) and Fig.\ \ref{fig:Bx}c), some modes with oscillations in the $O_x$ direction appear, which are absent in Fig.\ \ref{fig:Bx}a). This is an evidence of ESKHI coupling to electromagnetic waves, induced by the external magnetic field, which is predicted in our analyses in Sec.\ \ref{sec:num}.

Comparing Fig.\ \ref{fig:Bx}a and Fig.\ \ref{fig:Bx}b, we can see that the wavelength of the dominant mode in $B_z$ gets smaller in presence of the external magnetic field, which is also consistent to our conclusion in Sec.\ \ref{sec:num}. To give a quantitative verification, we carry out Fast Fourier Transform (FFT) to the transverse magnetic field $B_z$ in all runs. 
\begin{figure}[h]
\centering
\includegraphics[width=0.95 \columnwidth]{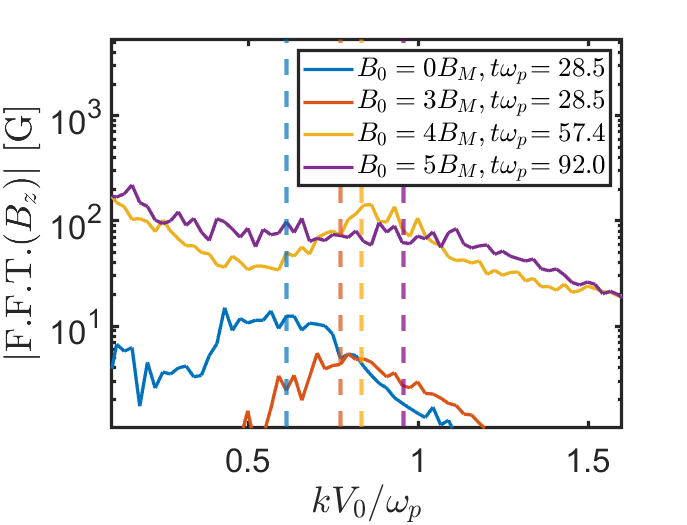}
\caption{The averaged amplitude of $B_z$'s FFT spectrum along the $O_y$ direction, marked in solid lines, in runs of $B_0=0,\ t=29.4\omega_p^{-1}$(blue), $B_0=3B_M,\ t=29.4\omega_p^{-1}$(red), $B_0=4B_M,\ t=59.2\omega_p^{-1}$(yellow), and $B_0=5B_M,\ t=59.2\omega_p^{-1}$(purple). The dashed vertical lines represents the wavenumbers $k_{\text{max}}$ with the maximum growth rate predicted for different magnitude of $B_0$ in Sec.\ \ref{sec:num}, where $k_{\text{max}}V_0/\omega_p=0.612$ (blue, 0$B_M$), 0.773 (red, 3$B_M$), 0.833 (yellow, 4$B_M$), and 0.958 (purple, 5$B_M$).}
\label{fig:fft}
\end{figure}
Figure\ \ref{fig:fft} shows the amplitude of $B_z$'s FFT spectrum along the $O_y$ direction, averaged in the $O_x$ direction. The wavenumbers $k_{\text{max}}$ with maximum growth rates predicted by our analyses in Sec.\ \ref{sec:num} is also shown for comparison. For each magnitude of the external magnetic field, the plotted time is chosen to be long enough for the modes to evolve but before non-linear effects dominate. The results of the simulations reach a fairly good agreement with the analyses, as the transformed $B_z$ shows a peak very close to the predicted wavenumber $k_{\text{max}}$ in each simulation, except when $B_0=5B_M$ and the mode drowned in non-linear behaviors before gets amplified significantly, due to a too small growth rate. The result of the FFT can be seen as a strong support to our theory and numerical solutions in Sec.\ \ref{sec:theo}.

\begin{figure}[h]
\centering
\includegraphics[width=0.95 \columnwidth]{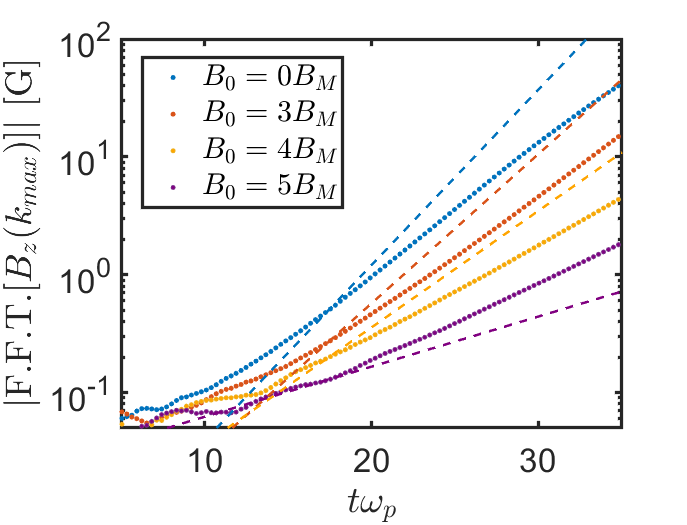}
\caption{The evolution of the averaged amplitude of $B_z$'s Fourier component, of the theoretically most unstable wavenumber $k_{\text{max}}$, marked in dots, in runs of $B_0=0$ (blue), $B_0=3B_M$ (red), $B_0=4B_M$ (yellow), and $B_0=5B_M$ (purple). The dashed lines represents exponential evolution with the theoretical maximum growth rates $\sigma_{\text{max}}$, for $B_0=0$ (blue), $B_0=3B_M$ (red), $B_0=4B_M$ (yellow), and $B_0=5B_M$ (purple). The scaling of vertical axis is logarithmic.}
\label{fig:fft2}
\end{figure}

We  would also like to verify the analytically calculated growth rates with the simulation results. To minimize the effects of the DC magnetic field, we still focus on the Fourier transform of $B_z$ and avoid an overall statistic of the total magnetic field. Figure\ \ref{fig:fft2} shows the evolution of the  $k_{\text{max}}$ component of the averaged amplitude of $B_z$'s FFT. Exponential evolution with the theoretical growth rates is also shown for comparison. From the figure, we can see that our theoretical prediction generally describes the trend of the growth rates in different simulations, while it does not match the exact magnitude of the growth rates. Beside the non-ideal effects, the discrepancy may imply the interplay between the DC magnetic field and ESKHI modes. For $B_0=0,3,4B_M$, the generation of DC magnetic field increases the magnetic pressure in the neighbourhood of the interface, pushing electrons out of the region, thus suppressing the growth of ESKHI modes. However, for $B_0=5B_M$ the growth rates of ESKHI modes are very small, and nonlinear evolution of the DC magnetic field may increase other components of the magnetic field on contrary.

\begin{figure}[h]
\centering
\includegraphics[width=0.95 \columnwidth]{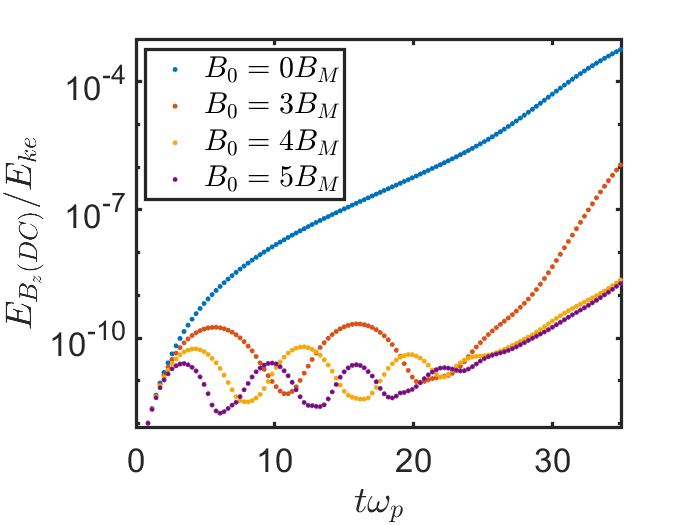}
\caption{The effective energy of the DC component $E_{B_z(\text{DC})}=\iint  \text{d}x\text{d}y\langle B_z \rangle_y^2/\mu_0$, divided by the initial kinetic energy of electrons, in runs of $B_0=0$ (blue), $B_0=3B_M$ (red), $B_0=4B_M$ (yellow), and $B_0=5B_M$ (purple). The scaling of vertical axis is logarithmic.}
\label{fig:dc}
\end{figure}

Speaking of the DC magnetic field, a remaining phenomenon in Fig.\ \ref{fig:Bx} is that the DC component of $B_z$ seems to be dominant in absence of the external magnetic field (Fig.\ \ref{fig:Bx}a) while dominated by the AC component in presence of the external magnetic field (Fig.\ \ref{fig:Bx}b). This indicates some mechanism suppressing the DC magnetic field in presence of the external magnetic field. Although the DC magnetic field is not the subject of this paper, it does not cost us extra work to reveal the cause of this phenomena. Figure\ \ref{fig:dc} shows the evolution of the DC component magnetic field in different runs. We average $B_z$ along the $Oy$ direction to derive the DC component $\langle B_z \rangle_y=\int \text{d}y B_z /L_y$, and sum over the simulation area to derive the effective energy $E_{B_z(\text{DC})}=\iint  \text{d}x\text{d}y\langle B_z \rangle_y^2/2\mu_0$.
 A clear difference is that $E_{B_z(\text{DC})}$ oscillates in the beginning of the simulations with external magnetic fields, while increases monotonically in the run without an external magnetic field. We should note that the generation of the DC magnetic field is induced by electron diffusion or mixing across the interface \citep{Grismayer2013dc}, where electrons initially locating in the layer with $v_y=V_0$ gain a $v_x$ across the interface and enter the other layer with $v_y=-V_0$ while the $y$ component of their velocity remains $v_y\approx V_0$, and vise versa. As the ions remain unperturbed, this mixing causes a current imbalance in the neighbourhood of the interface and results in the generation of DC magnetic field. Bearing this in mind, we can easily point out that the oscillation of the DC field energy is caused by electron gyration in the $xOz$ plane, where the first electrons crossing the interface re-enter their initial layer, mitigating the current imbalance, and this circulation repeats. From Fig.\ \ref{fig:dc} one can see that the oscillation frequency of the DC field is just positively related with the electron cyclotron frequency $\omega_c$. After about $t=25\omega_p^{-1}$, the gyrations are overwhelmed by the net diffusion across the interface, and the DC field begins to increase monotonically.

Our analyses of the DC magnetic field are very simple and qualitative. An accurate description of the DC magnetic field requires a kinetic treatment of the electron distribution function \citep{Grismayer2013dc}, which is beyond the subject of this work. Also, the interplay between the DC magnetic field and ESKHI modes is rather unknown. As the generation of the DC field is inevitable in typical ESKHI settings and its magnitude comparable with the AC field induced by ESKHI, it requires a description of the interplay between them to fully understand the behavior of relativistic shear flows.

\section{Conclusion} \label{sec:conclusion}

In this paper, we carried out an analytical study and also simulations on ESKHI under external magnetic fields. In the limit of a cold and collisionless plasma, for the first time, we derived the generalized eigenvalue equations Eqs.\ \ref{eq:dis1} and \ref{eq:dis2} for magnetized ESKHI in the relativistic regime. We numerically solved the dispersion relation of ESKHI growth rates for some simple cases, where the shear flow is symmetrical and the external magnetic field is uniform or anti-parallel. We found that the unstable modes of ESKHI survives beyond the threshold magnetic field $B_M$ proposed by MHD theory, and the true threshold field can be much larger. Also, uniform and anti-parallel external magnetic fields result in different dispersion relations, which is different from the conclusion of the MHD theory. The existence of an external magnetic field decreases the maximum growth rate of ESKHI for most shear velocities, but increases it for a sufficiently large shear velocity. The external field also activates modes which are stable in absence of the magnetic field, and results in a new normalized cutoff wavenumber dependent on only $\omega_c$ and $\omega_p$. For $V_0=0.2c$, as the magnitude of the external magnetic field increases, the maximum growth rate gets smaller while the corresponding wavenumber gets larger, and both conclusions are verified by the PIC simulations. For some wavenumbers, the external magnetic field induces coupling with electromagnetic waves, which results in modes with complex eigenfrequencies. Moreover, in PIC simulations we find that the external magnetic field suppresses the generation of the DC magnetic field, which is directly related to electron gyration in the neighbourhood of the shear interface.

Although our generalized eigenvalue equations for ESKHI are explicit, the mathematical details of the equations and corresponding physical behaviors are still not fully explored. Cases with different profiles of zeroth-order variables may be tested. Future studies along with simulations might verify and explain the anomalous destabilizing for large shear velocities, the coupling with electromagnetic waves, and the activation of perturbations with larger wavenumbers. Exact form of the cutoff wavenumber, threshold magnetic field and the turning point might also be derived with some delicate mathematical treatment.

Moreover, our theoretical analyses were carried out in the limit of cold plasma. Extension to the finite-temperature regime can be done by including the electron pressure terms in the relativistic fluid equations \citep{miller2016relativistic}. Once extended to the finite-temperature regime, the equations can be used to study equilibrium states with any profile of density and magnetic field.  Also, we only considered perturbations alongside the streaming direction, while perturbations along the transverse direction will give rise to a different instability MI \citep{Alves2015Transverse}. A once-for-all dispersion relation encompassing ESKHI, MI, and their coupling could be done by considering perturbations with the form of $f(x)e^{i(k_yy+k_zz-\omega t)}$, which is a direct extension to our work.

Finally, our theoretical analyses were based on linearized fluid equations and did not contain the generation of the DC magnetic field, which has to be determined in a kinetic approach \citep{Grismayer2013dc}. Since the generation of DC magnetic fields is inevitable in the ESKHI settings and must have an interplay with ESKHI, understanding it is necessary to break down the magnetic field generation mechanism in relativistic shear flows, and might be the motivation of future works.

%% IMPORTANT! The old "\acknowledgment" command has be depreciated. It was
%% not robust enough to handle our new dual anonymous review requirements and
%% thus been replaced with the acknowledgment environment. If you try to 
%% compile with \acknowledgment you will get an error print to the screen
%% and in the compiled pdf.
%% 
%% Also note that the akcnowlodgment environment does not support long amounts of text. If you have a lot of people and institutions to acknowledge, do not use this command. Instead, create a new \section{Acknowledgments}.
\begin{acknowledgments}
Yao Guo would like to thank Dr.\ Tian-Xing Hu for helpful discussions. This work was supported by the Strategic Priority Research Program of Chinese Academy of Sciences (Grant No. XDA250010100 and XDA250050500), National Natural Science Foundation of China (Grant No. 12075204) and Shanghai Municipal Science and Technology Key Project (Grant No. 22JC1401500). Dong Wu thanks the sponsorship from Yangyang Development Fund.
\end{acknowledgments}

%% To help institutions obtain information on the effectiveness of their 
%% telescopes the AAS Journals has created a group of keywords for telescope 
%% facilities.
%
%% Following the acknowledgments section, use the following syntax and the
%% \facility{} or \facilities{} macros to list the keywords of facilities used 
%% in the research for the paper.  Each keyword is check against the master 
%% list during copy editing.  Individual instruments can be provided in 
%% parentheses, after the keyword, but they are not verified.

%\vspace{5mm}
%\facilities{HST(STIS), Swift(XRT and UVOT), AAVSO, CTIO:1.3m, CTIO:1.5m,CXO}

%% Similar to \facility{}, there is the optional \software command to allow 
%% authors a place to specify which programs were used during the creation of 
%% the manuscript. Authors should list each code and include either a
%% citation or url to the code inside ()s when available.

%\software{astropy \citep{2013A&A...558A..33A,2018AJ....156..123A},  
%          Cloudy \citep{2013RMxAA..49..137F}, 
%          Source Extractor \citep{1996A&AS..117..393B}
%          }

%% Appendix material should be preceded with a single \appendix command.
%% There should be a \section command for each appendix. Mark appendix
%% subsections with the same markup you use in the main body of the paper.

%% Each Appendix (indicated with \section) will be lettered A, B, C, etc.
%% The equation counter will reset when it encounters the \appendix
%% command and will number appendix equations (A1), (A2), etc. The
%% Figure and Table counter will not reset.

\appendix

\section{Numerical scheme for solving the generalized eigenvalue equation}\label{sec:app}

Here we introduce a scheme to numerically solve the generalized eigenvalue equations Eqs.\ \ref{eq:dis1} and \ref{eq:dis2}. In the case of semi-uniform shear flows (the zeroth-order variables $n_0,V_0,B_0$ are uniform at least on each half of the plane), Eqs.\ \ref{eq:dis1}  and \ref{eq:dis2} can be decoupled, resulting in two four-order differential equations, and admits symbolic solutions to $E_y(x)$ and $E_z(x)$.  Although the explicit form of the solution is rather enormous, one can at least acknowledge that the general structure of the solution to $E_y(x)$ and $E_z(x)$ is 
\begin{align}
    E_y^\pm(x)=f_1^\pm(\omega)\exp[g_1^\pm(\omega)x]+f_2^\pm(\omega)\exp[-g_1^\pm(\omega)x]+f_3^\pm(\omega)\exp[g_2^\pm(\omega)x]+f_4^\pm(\omega)\exp[-g_2^\pm(\omega)x],\label{eq:sol1} \\ 
    E_z^\pm(x)=h_1^\pm(\omega)\exp[g_1^\pm(\omega)x]+h_2^\pm(\omega)\exp[-g_1^\pm(\omega)x]+h_3^\pm(\omega)\exp[g_2^\pm(\omega)x]+h_4^\pm(\omega)\exp[-g_2^\pm(\omega)x], \label{eq:sol2}
\end{align}
 on each half of the plane, with the $+$ superscript for $x>0$ and $-$ for $x<0$. This results in 10 unknown functions on each side of the plane and 21 unknowns in total (including the undetermined eigenfrequency $\omega$). 
 
 Now we take a shear flow in the infinite plane as an example. Without loss of generality, we can assume $\text{Re}[g^\pm_i(\omega)]>0$. Modes divergent at infinities should be eliminated, i.e. $f_1^+,f_3^+,h_1^+,h_3^+,f_2^-,f_4^-,h_2^-,$ and $h_4^-=0$. This reduces 8 unknowns from the system, and only 2 linear independent modes remain on each half of the plane (exp$[-g^+_{1,2}(\omega)x]$ for the upper plane and exp$[g^-_{1,2}(\omega)x]$ for the lower plane). Then we substitute the solution Eqs.\ \ref{eq:sol1} and \ref{eq:sol2} into Eqs.\ \ref{eq:dis1} and \ref{eq:dis2} on each half of the plane. It is clear that each of the remaining 4 independent modes in the solution should satisfy Eqs.\ \ref{eq:dis1} and \ref{eq:dis2} respectively. That is, for the mode $\text{exp}[-g^+_{1}(\omega)x]$, $E_y^+(x)=f_2^+(\omega)\text{exp}[-g^+_{1}(\omega)x]$ together with $E_z^+(x)=h_2^+(\omega)\text{exp}[-g^+_{1}(\omega)x]$ should satisfy Eqs.\ \ref{eq:dis1} and \ref{eq:dis2}, and so do other 3 modes. This results in 8 polynomial equations in total. Equations \ref{eq:dis10} and \ref{eq:dis20}
 along with the electric field continuous conditions $E_y(0_+)=E_y(0_-)$ and $E_z(0_+)=E_z(0_-)$ give another 4 equations. Finally, since the amplitude of the perturbations can be arbitrary, one can normalize the coefficient of any one of the remaining modes to be a certain value. Now 13 polynomial equations are given for 13 unknowns and the equation is numerically solvable. This method can also be applied to shear flows with other boundary conditions.
 
For shears with smooth velocity or density profiles (e.g. $V_0(x)=\text{tanh}(x)V_0$), one can just use the traditional shooting method to solve the eigenfrequencies. We should note that the solutions to $E_y(x)$ and $E_z(x)$ contain $\omega$ itself, which means we can not find an accurate boundary condition \textit{apriori} and iterations are needed. We can start with solutions derived for the uniform case as boundary conditions at one finite end, and numerically solve Eqs.\ \ref{eq:dis1} and \ref{eq:dis2} with different trial values of $\omega$. We search for the best $\omega$ to make $E_y(x)$ and $E_z(x)$ convergent at the other end. This new $\omega$ results in a new solution to $E_y(x)$ and $E_z(x)$, which in turn can be used as the new boundary conditions. Repeating this process until $\omega$ converges gives the final solution of $\omega$. It is better to start with profiles with larger gradients, and use their $\omega$ as the starting trial value for profiles with smaller gradients.

\bibliography{sample631}{}

\begin{thebibliography}{}
\expandafter\ifx\csname natexlab\endcsname\relax\def\natexlab#1{#1}\fi
\providecommand{\url}[1]{\href{#1}{#1}}
\providecommand{\dodoi}[1]{doi:~\href{http://doi.org/#1}{\nolinkurl{#1}}}
\providecommand{\doeprint}[1]{\href{http://ascl.net/#1}{\nolinkurl{http://ascl.net/#1}}}
\providecommand{\doarXiv}[1]{\href{https://arxiv.org/abs/#1}{\nolinkurl{https://arxiv.org/abs/#1}}}

\bibitem[{Alves {et~al.}(2014)Alves, Grismayer, Fonseca, \&
  Silva}]{Alves2014electronshear}
Alves, E.~P., Grismayer, T., Fonseca, R.~A., \& Silva, L.~O. 2014, New Journal
  of Physics, 16, 035007, \dodoi{10.1088/1367-2630/16/3/035007}

\bibitem[{Alves {et~al.}(2015)Alves, Grismayer, Fonseca, \&
  Silva}]{Alves2015Transverse}
---. 2015, Phys. Rev. E, 92, 021101, \dodoi{10.1103/PhysRevE.92.021101}

\bibitem[{Alves {et~al.}(2012)Alves, Grismayer, Martins, Fiúza, Fonseca, \&
  Silva}]{Alves2012large}
Alves, E.~P., Grismayer, T., Martins, S.~F., {et~al.} 2012, The Astrophysical
  Journal Letters, 746, L14, \dodoi{10.1088/2041-8205/746/2/L14}

\bibitem[{B{\"o}ttcher(2007)}]{bottcher2007modeling}
B{\"o}ttcher, M. 2007, 95, \dodoi{10.1007/s10509-007-9404-0}

\bibitem[{Chandrasekhar(1961)}]{chandrasekhar1961hydrodynamic}
Chandrasekhar, S. 1961, Hydrodynamic and hydromagnetic stability (Clarendon
  Press)

\bibitem[{Che \& Zank(2023)}]{Che2023EKHI}
Che, H., \& Zank, G.~P. 2023, Physics of Plasmas, 30, 062110,
  \dodoi{10.1063/5.0150895}

\bibitem[{Colgate {et~al.}(2001)Colgate, Li, \& Pariev}]{Colgate2001}
Colgate, S.~A., Li, H., \& Pariev, V. 2001, Physics of Plasmas, 8, 2425,
  \dodoi{10.1063/1.1351827}

\bibitem[{Coppi {et~al.}(1966)Coppi, Laval, \& Pellat}]{Coppi1966}
Coppi, B., Laval, G., \& Pellat, R. 1966, Phys. Rev. Lett., 16, 1207,
  \dodoi{10.1103/PhysRevLett.16.1207}

\bibitem[{D'Angelo(1965)}]{Dangelo1965}
D'Angelo, N. 1965, The Physics of Fluids, 8, 1748, \dodoi{10.1063/1.1761496}

\bibitem[{Fermo {et~al.}(2012)Fermo, Drake, \& Swisdak}]{Fermo2012Secondary}
Fermo, R.~L., Drake, J.~F., \& Swisdak, M. 2012, Phys. Rev. Lett., 108, 255005,
  \dodoi{10.1103/PhysRevLett.108.255005}

\bibitem[{Grismayer {et~al.}(2013)Grismayer, Alves, Fonseca, \&
  Silva}]{Grismayer2013dc}
Grismayer, T., Alves, E.~P., Fonseca, R.~A., \& Silva, L.~O. 2013, Phys. Rev.
  Lett., 111, 015005, \dodoi{10.1103/PhysRevLett.111.015005}

\bibitem[{Gruzinov(2008)}]{Gruzinov2008grb}
Gruzinov, A. 2008, arXiv preprint arXiv:0803.1182.
\newblock \url{https://doi.org/10.48550/arXiv.0803.1182}

\bibitem[{Gruzinov \& Waxman(1999)}]{Gruzinov1999Gamma}
Gruzinov, A., \& Waxman, E. 1999, The Astrophysical Journal, 511, 852,
  \dodoi{10.1086/306720}

\bibitem[{Helmholtz(1868)}]{helmholtz1868}
Helmholtz. 1868, The London, Edinburgh, and Dublin Philosophical Magazine and
  Journal of Science, 36, 337

\bibitem[{Jiang {et~al.}(2025)Jiang, Huang, Lu, Yuan, \& Xiong}]{Jiang2025}
Jiang, K., Huang, S.~Y., Lu, Q.~M., Yuan, Z.~G., \& Xiong, Q.~Y. 2025,
  Geophysical Research Letters, 52, e2024GL111450,
  \dodoi{https://doi.org/10.1029/2024GL111450}

\bibitem[{Kelvin(1871)}]{Kelvin1871}
Kelvin. 1871, The London, Edinburgh, and Dublin Philosophical Magazine and
  Journal of Science, 42, 362, \dodoi{10.1080/14786447108640585}

\bibitem[{Liang {et~al.}(2013{\natexlab{a}})Liang, Boettcher, \&
  Smith}]{Liang2013Magnetic}
Liang, E., Boettcher, M., \& Smith, I. 2013{\natexlab{a}}, The Astrophysical
  Journal Letters, 766, L19, \dodoi{10.1088/2041-8205/766/2/L19}

\bibitem[{Liang {et~al.}(2013{\natexlab{b}})Liang, Fu, Boettcher, Smith, \&
  Roustazadeh}]{Liang2013Relativistic}
Liang, E., Fu, W., Boettcher, M., Smith, I., \& Roustazadeh, P.
  2013{\natexlab{b}}, The Astrophysical Journal Letters, 779, L27,
  \dodoi{10.1088/2041-8205/779/2/L27}

\bibitem[{Liang {et~al.}(2017)Liang, Fu, \& Böttcher}]{Liang_2017}
Liang, E., Fu, W., \& Böttcher, M. 2017, The Astrophysical Journal, 847, 90,
  \dodoi{10.3847/1538-4357/aa8772}

\bibitem[{Mandelker {et~al.}(2016)Mandelker, Padnos, Dekel, Birnboim, Burkert,
  Krumholz, \& Steinberg}]{Mandelker2016}
Mandelker, N., Padnos, D., Dekel, A., {et~al.} 2016, Monthly Notices of the
  Royal Astronomical Society, 463, 3921, \dodoi{10.1093/mnras/stw2267}

\bibitem[{Medvedev \& Loeb(1999)}]{Medvedev1999}
Medvedev, M.~V., \& Loeb, A. 1999, The Astrophysical Journal, 526, 697,
  \dodoi{10.1086/308038}

\bibitem[{Miller \& Rogers(2016)}]{miller2016relativistic}
Miller, E.~D., \& Rogers, B.~N. 2016, Journal of Plasma Physics, 82, 905820205,
  \dodoi{10.1017/S0022377816000180}

\bibitem[{Mirabel \& Rodríguez(1999)}]{Mirabel1999}
Mirabel, I.~F., \& Rodríguez, L.~F. 1999, Annual Review of Astronomy and
  Astrophysics, 37, 409, \dodoi{https://doi.org/10.1146/annurev.astro.37.1.409}

\bibitem[{Niu {et~al.}(2025)Niu, Wang, Fu, Cao, \& Fu}]{Niu2025}
Niu, J.~Q., Wang, Z., Fu, H.~S., Cao, J.~B., \& Fu, W.~D. 2025, Journal of
  Geophysical Research: Space Physics, 130, e2024JA033473,
  \dodoi{https://doi.org/10.1029/2024JA033473}

\bibitem[{Ohira(2013)}]{Ohira2013}
Ohira, Y. 2013, The Astrophysical Journal Letters, 767, L16,
  \dodoi{10.1088/2041-8205/767/1/L16}

\bibitem[{Piran(2005)}]{Piran2005}
Piran, T. 2005, Rev. Mod. Phys., 76, 1143, \dodoi{10.1103/RevModPhys.76.1143}

\bibitem[{Pritchett {et~al.}(1996)Pritchett, Coroniti, \&
  Decyk}]{Pritchett1996}
Pritchett, P.~L., Coroniti, F.~V., \& Decyk, V.~K. 1996, Journal of Geophysical
  Research: Space Physics, 101, 27413,
  \dodoi{https://doi.org/10.1029/96JA02665}

\bibitem[{Rieger(2019)}]{Rieger2019}
Rieger, F.~M. 2019, Galaxies, 7, \dodoi{10.3390/galaxies7030078}

\bibitem[{Rieger \& Duffy(2004)}]{Rieger2004}
Rieger, F.~M., \& Duffy, P. 2004, The Astrophysical Journal, 617, 155,
  \dodoi{10.1086/425167}

\bibitem[{Silva {et~al.}(2003)Silva, Fonseca, Tonge, Dawson, Mori, \&
  Medvedev}]{Silva_2003}
Silva, L.~O., Fonseca, R.~A., Tonge, J.~W., {et~al.} 2003, The Astrophysical
  Journal, 596, L121, \dodoi{10.1086/379156}

\bibitem[{Tsiklauri(2024)}]{Tsiklauri_2024}
Tsiklauri, D. 2024, Research in Astronomy and Astrophysics, 24, 095021,
  \dodoi{10.1088/1674-4527/ad74de}

\bibitem[{Weibel(1959)}]{Weibel1959}
Weibel, E.~S. 1959, Phys. Rev. Lett., 2, 83, \dodoi{10.1103/PhysRevLett.2.83}

\bibitem[{Wu {et~al.}(2019)Wu, Yu, Fritzsche, \& He}]{Wu2019}
Wu, D., Yu, W., Fritzsche, S., \& He, X.~T. 2019, Phys. Rev. E, 100, 013207,
  \dodoi{10.1103/PhysRevE.100.013207}

\bibitem[{Yao {et~al.}(2020)Yao, Cai, Yan, Zhang, Du, Tian, Zhang, Wang, \&
  Zhu}]{Yao2020}
Yao, P., Cai, H., Yan, X., {et~al.} 2020, Matter and Radiation at Extremes, 5,
  054403, \dodoi{10.1063/5.0017962}

\bibitem[{Zhong {et~al.}(2018)Zhong, Tang, Zhou, Deng, Pang, Paterson, Giles,
  Burch, Tobert, Ergun, Khotyaintsev, \& Lindquist}]{Zhong2018Evidence}
Zhong, Z.~H., Tang, R.~X., Zhou, M., {et~al.} 2018, Phys. Rev. Lett., 120,
  075101, \dodoi{10.1103/PhysRevLett.120.075101}

\end{thebibliography}
\bibliographystyle{aasjournal}

%% This command is needed to show the entire author+affiliation list when
%% the collaboration and author truncation commands are used.  It has to
%% go at the end of the manuscript.
%\allauthors

%% Include this line if you are using the \added, \replaced, \deleted
%% commands to see a summary list of all changes at the end of the article.
%\listofchanges

\end{document}